\documentstyle[twocolumn,prl,floats,aps,psfig]{revtex}
\newcommand{\be}{\begin{equation}}
\newcommand{\ee}{\end{equation}}

\begin{document}
\draft

%
%
\twocolumn[\hsize\textwidth\columnwidth\hsize\csname @twocolumnfalse\endcsname

\title{ Hole Dispersion and Symmetry
of the Superconducting Order Parameter for Underdoped CuO$_2$
Bilayers and 3D Antiferromagnets.}

\author{ Alexander Nazarenko and Elbio Dagotto}
\address{Department of Physics and National High Magnetic Field Lab,
Florida State University, Tallahassee, FL 32306, USA}

\maketitle

\begin{abstract}
\noindent
We calculate the dispersion of a hole in
${\rm CuO_2}$ bilayers and 3D antiferromagnets, using the
self-consistent Born approximation (SCBA)
to the $t-J$ model. Superconductivity and the symmetry of its
order parameter are studied
introducing a nearest-neighbor density-density 
attraction induced by short range antiferromagnetic (AF)
fluctuations, as described in recent studies for the single layer cuprates.
The well-known pairing in 
the ${ d_{x^2-y^2}}$  channel  observed for one plane remains
 robust when three dimensional interactions are turned on.
In bilayers, as the exchange along the direction 
perpendicular to the planes 
grows, eventually a transition 
to a ``s-wave'' state is observed with an order
parameter mixture of
${ d_{3z^2-r^2}}$ and ${ s_{x^2+y^2 + z^2}}$.
For an isotropic 3D antiferromagnet the $d_{x^2 -y^2}$ and $d_{3z^2 - r^2}$
channels are degenerate.
Our results are compared with other predictions in the literature.


\end{abstract}

]
%
%

\section{Introduction}

  Since the discovery of high temperature superconductors, the origin
of the pairing mechanism has been controversial. Numerous studies
suggest that the normal state properties are anomalous and, as a consequence,
above the critical temperature, ${ T_c}$, the strongly correlated
character of the conduction electrons in the cuprates should be taken
into account in any realistic study. In particular, the presence of
short-range antiferromagnetic fluctuations have been invoked by several
groups as the origin of the deviations  from a conventional Fermi liquid
behavior observed in transport
measurements above $T_c$. Theories based on antiferromagnetic
fluctuations quite clearly predict a superconducting condensate
in the $d_{x^2-y^2}$ channel.\cite{AF-potential,review} 
These studies are supported by
experimental results for the bilayer cuprate
${\rm YBa_2Cu_3O_{7-\delta}}$, which are compatible with
a highly anisotropic pairing state, likely a ${ d_{x^2-y^2}}$
singlet.\cite{gap-symmetry}
Photoemission results for Bi-2212, which also
have two $CuO_2$ layers per unit cell, support this
scenario,\cite{ARPES} and thus  
superconductivity induced by AF correlations is a strong candidate to explain
the pairing mechanism of the cuprates.

The authors and collaborators
have recently explored these ideas using simple models for
quasiparticles (q.p.) considered as holes
strongly dressed by AF fluctuations, and interacting
through a nearest-neighbor (NN) density-density attraction, as suggested
by the study of
the two-holes bound-state of the $t-J$ model.\cite{afvh} 
The presence of a large
accumulation of weight in the density of states, induced by the small
bandwidth of the q.p. dispersion, enhances $T_c$ beyond what
would be naively expected from the strength of the hole-hole attraction.
These ideas were originally referred to as the ``antiferromagnetic
van Hove'' (AFVH) scenario, but it is important to remark that the boost
in $T_c$ does not arise exclusively from a van Hove singularity but
mainly from the ``flat'' regions observed in the q.p. dispersion, effect
which is induced by AF correlations.\cite{flat-theory,other-flat}

The purpose of this paper is (i) the study of the properties of holes and
(ii) the extension of  previous analysis for 
the symmetry of the
superconducting order parameter to the case of  lightly doped bilayers and 3D 
antiferromagnets.
%
%
The effective Hamiltonian introduced in previous 
publications for one plane\cite{afvh} is defined on the 2D
square lattice as
$$
H = \sum_{{\bf k}\alpha} \epsilon_{AF}({\bf k}) 
c^\dagger_{{\bf k}\alpha} c_{{\bf
k}\alpha} 
- V \sum_{\langle {\bf ij} \rangle } n_{\bf i} n_{\bf j},
\eqno(1)
$$
where $c_{{\bf k}\alpha}$ denotes a destruction operator of a q.p. 
with dispersion
extracted from accurate numerical and analytical studies of one hole in
the undoped $t-J$ model given by
 $\epsilon_{AF}({\bf k})/eV=0.165\cos k_x\cos k_y\,
+\,0.0435(\cos 2k_x+\cos 2k_y)$, where $t=0.4eV$ and $J/t = 0.3$ are
assumed.\cite{flat-theory} 
The
quasiparticles move within the same sublattice to avoid distorting the
AF background ($\alpha=A,B$ denotes the sublattice). 
This is correct in the limit of a small number of holes
in a long-range ordered AF system, and it is expected to be a good
approximation even with AF short-range order, as long as the AF 
correlations are strong.
The dispersion $\epsilon_{AF}({\bf k})$
reflects a remarkable feature of the cuprates, observed in photoemission
experiments, namely the presence of flat bands near $(0,\pi)$ $(\pi,0)$
in the standard square lattice notation.
\cite{flat-theory,other-flat} 
The parameter $V$ is the intersublattice density-density
attractive  coupling strength between holes also
discussed in previous literature,\cite{afvh} with
$n_{\bf i}$ the number operator. 
The rest of the notation is standard.

The interaction in Eq.(1) should be
considered as the dominant piece  of a more general and extended 
AF-induced effective potential
between holes which in the presence of long-range order
corresponds to a sharp $\delta$-function 
centered at ${\bf
Q}=(\pi,\pi)$ in momentum space, and it 
acquires a width as the AF
correlation length $\xi_{AF}$ decreases. Even with $\xi_{AF}$ as small
as a couple of lattice spacings, it has been shown that the 
NN interaction used in Eq.(1) remains robust.\cite{dos}
Equivalently, the NN hole-hole
attraction can be considered as arising from the ``minimization of
broken AF links'' argument\cite{review} in the large $J/t$ limit.


It is expected that
Hamiltonian Eq.(1), which is certainly a very simplified version of the
low energy behavior of the $t-J$ model, nevertheless captures
the important features of holes moving in AF backgrounds at low doping,
and temperatures smaller than $J$ where the quasiparticles are dominant.
This is a regime difficult to study with numerical methods directly applied
to the $t-J$ model.
The analysis of Eq.(1) using standard BCS techniques, Exact
Diagonalization approaches,
and the Eliashberg equations, has been fairly 
successful in reproducing some experimental 
features of the cuprates.\cite{afvh,naza1}
For example, in this model 
superconductivity appears in the $d_{x^2-y^2}$ channel, as it seems to
occur in
experiments,
 with a critical temperature $T_c \sim 100 K$. In addition,  the concept of 
``optimal doping'' appears naturally due to the robust peak in the 
density of states (DOS) produced by the dispersion $\epsilon_{AF}({\bf
k})$,
 which is crossed by the chemical potential as
the hole density grows. In this respect, the theory has features very
similar to those discussed before in ``van Hove'' theories for the
cuprates.\cite{vh} Indeed, the quasiparticle lifetime 
is linear with energy at optimal doping. 
However, the flat regions in the dispersion and the associated
accumulation of weight in the DOS are produced by strong
correlations and thus they are
stable against small perturbations, like impurities and extra couplings,
effects that usually destroy  weak logarithmic van Hove singularities.
The specific heat jump, 
$2\Delta/T_c$ and other important
BCS ratios are in good agreement with the experimental data.\cite{afvh}

Although certainly more work is needed to show that
these ``real-space'' pairing ideas presented in previous literature are
a strong candidate to describe the cuprates, its
quantitative success led us here to study geometries and couplings beyond
those of the single layer with some confidence.
Our goal is to report the trends observed when systems different
from a single layer cuprate are analyzed.
%
%
Here, the hole spectrum in the bilayer and 3D antiferromagnets is
calculated with the SCBA
which was shown to reproduce accurately 
numerical results for the 2D $t-J$ model,\cite{Rainbow}
and it was successfully applied to
the 2D $t-t^{\prime}-J$ model to compare its predictions with the photoemission
spectra of ${\rm Sr_2CuO_2Cl_2}$.\cite{Nazarenko} Using the formalism
previously
described, a hole-hole attraction will be introduced
producing a superconducting state. Here, the symmetry of
this superconducting state will be analyzed
and compare with other predictions for the
same bilayer and 3D systems. 


\section{Self Consistent Born Approximation.}

The Hamiltonian for spins and holes used in this section is defined as,
$$ H  =  
-\sum_{\alpha} t_{\alpha} \sum_{ {\bf i}\sigma }
({\bar c}^{\dagger}_{{\bf i}\sigma} 
{\bar c}_{{\bf i}+{\bf e}_{\alpha}\sigma} +h.c.) 
$$
$$
-\frac{1}{2} {\textstyle {\displaystyle\sum_{ \alpha\beta }}}
t_{ \alpha\beta } \sum_{ {\bf i}\sigma } (
{\bar c}^{\dagger}_{{\bf i}\sigma} 
{\bar c}_{{\bf i}+{\bf e}_{\alpha}\pm{\bf e}_{\beta}\sigma} + h.c.)~~ $$ 
$$ + \sum_{\alpha}J_{\alpha}\sum_{ {\bf i} }
[ ( S^{z}_{{\bf i}} S^{z}_{{\bf i}+{\bf e}_{\alpha}} - 
{{1}\over{4}} n_{\bf i} n_{{\bf i}+{\bf e}_{\alpha}})~~ 
$$
$$
+ \frac{1}{2} \vartheta_{\alpha}
( S^{+}_{{\bf i}}   S^{-}_{{\bf i}+{\bf e}_{\alpha}}+
  S^{-}_{{\bf i}}   S^{+}_{{\bf i}+{\bf e}_{\alpha}}) ],  
\eqno(2)
$$
\noindent
where ${\bf i}$ denote sites of a bilayer or simple 3D cubic cluster,
${\bf e}_{\alpha}$ is the unit vector in the $\alpha$-direction 
($\alpha=x,y,z$),
$t_{\alpha}$ and $J_{\alpha}$ correspond to the NN 
hopping amplitude and exchange coupling, respectively, 
in the direction $\alpha$,
$t_{ \alpha\beta }$ is the next-nearest-neighbors 
(NNN) hopping in the directions defined by ${\bf
e}_{\alpha} \pm {\bf e}_{\beta}$,
the parameter
$\vartheta_{\alpha} ( \epsilon [0,1] )$ represents a 
possible exchange anisotropy added for
completeness, and the rest of
the notation is standard.
In the terms with NNN interactions, the summation
is done with the condition $\alpha\neq\beta$, i.e. we only
consider NNN hopping along the diagonals of the plaquettes. A
generalization to include NNN hopping at distance of two lattice
spacings along the main axes is straightforward.
The constraint of no double occupancy is implemented in the
kinetic energy by means of the  standard definition for hole operators 
${\bar c}_{{\bf i}\sigma}~=~c_{{\bf i}\sigma}(1-n_{{\bf i}-\sigma})$, with
$n_{{\bf i}\sigma}=c^{\dagger}_{{\bf i}\sigma}c_{{\bf i}\sigma}$.

Let us extend the SCBA analysis which has been 
applied before to single planes\cite{Rainbow} to the bilayer and 3D
Hamiltonian Eq.(2). Following steps that by now are standard, first
we redefine operators according to  $S_{\bf j}^{\pm}
\rightarrow S_{\bf j}^{\mp}$, $S_{\bf j}^{z}
\rightarrow -S_{\bf j}^{z}$, $c_{{\bf j}\sigma}\rightarrow
c_{{\bf j}-\sigma}$, where ${\bf j}$ denotes sites on the $B$
sublattice. Spins on the $A$ sublattice remain intact.
With this  procedure an AF state becomes ferromagnetic.
Then, the usual linearized Holstein-Primakoff transformation is used. This
transformation is defined in terms
of Bose operators $a_{\bf i}$ and $a^{\dagger}_{{\bf i}}$, according to
$S_{{\bf i}}^+ \approx a_{{\bf i}}\sqrt {2S}$,
$S_{{\bf i}}^- \approx a^{\dagger}_{{\bf i}}\sqrt {2S}$, and
$S_{{\bf i}}^z=S-a^{\dagger}_{{\bf i}}a_{{\bf i}}$. To handle the hole
sector, the 
spin-charge decomposition $c_{{\bf i}\uparrow}=h^{\dagger}_{{\bf i}}$
and
$c_{{\bf i}\downarrow} = h^{\dagger}_{{\bf i}}S_{{\bf i}}^+$ is 
introduced, where
$h_{\bf i},h^{\dagger}_{\bf i}$ are operators corresponding to spinless holes.
Finally, using periodic boundary conditions and after long but
straightforward algebra,
we can rewrite the resulting Hamiltonian in momentum space as
$$
H = E_0 + \sum_{\bf k}\epsilon_{\bf k}h^{\dagger}_{\bf k}h_{\bf k} +
\sum_{\bf q}\omega_{\bf q}b^{\dagger}_{\bf q}b_{\bf q}$$
$$+ \frac{1}{\sqrt N}\sum_{{\bf k}{\bf q}}[
M_{{\bf k},{\bf q}}h^{\dagger}_{\bf k}h_{{\bf k}-{\bf q}}b_{\bf q}
+ h.c.],
\eqno(3)
$$
\noindent
where the Bose operators
$b(b^{\dagger})$ and $a(a^{\dagger})$ are
connected by a Bogoliubov transformation,\cite{Rainbow}
$N$ is the number of sites in the lattice, 
$\epsilon_{\bf k}=4(t_{xy}\cos k_x\cos k_y
+t_{xz}\cos k_x\cos k_z+t_{yz}\cos k_y\cos k_z)$
is the bare spinless
hole dispersion (which cancels in the absence of an explicit
NNN hopping  in the Hamiltonian), and
$M_{{\bf k},{\bf q}} = u_{\bf q}\beta_{{\bf k}-{\bf q}}+
v_{\bf q}\beta_{\bf k}$ is the hole-magnon vertex. The magnon dispersion
is given by
$\omega_{\bf q}=\omega_0\nu_{\bf q}$, where
$\omega_0=zSJ$, $J=\frac{1}{d}
\sum_{\alpha}J_{\alpha}$,  
$S=\frac{1}{2}$ is the spin of the electrons in the original $t-J$ model
language,
$d$ is the number of dimensions (which will be taken equal to 3 in the
rest of the paper), 
$z=2d$ is the coordination number, and we have used units where $\hbar=1$.
The other quantities are defined as follows:
$$
\beta_{\bf k} = 2
\sum_{\alpha}
t_{\alpha}\cos k_{\alpha},
$$
$$
u_{\bf q}=\sqrt \frac{1+\nu_{\bf q}}{2\nu_{\bf q}},\,\,\,\,\,
v_{\bf q}=-sgn \kappa_{\bf q}\sqrt \frac{1-\nu_{\bf q}}{2\nu_{\bf q}},
$$
$$
\nu_{\bf q}=\sqrt {1-\kappa^{2}_{\bf q}},\,\,\,\,\,
\kappa_{\bf q}= \frac{
2S\sum_{{\alpha}}
J_{\alpha}\vartheta_{\alpha}\cos q_{\alpha}}
{\omega_0}.
\eqno(4)$$
\noindent
The ground
state energy $E_0$ has the form:
$$
E_0=-dJN[S^2+S(1-\frac{1}{N}\sum_{\bf q}\nu_{\bf q})+\frac{1}{4}].
\eqno(5)
$$

Care must be taken in order to apply Eqs.(3) and (4) to the bilayer. The
small size in the z-direction (one lattice spacing)
does not allow us to apply periodic
boundary conditions (BC) to simulate bulk effects, as we do in the x- and
y-directions. The z-axis BC should be open. However, for a bilayer
open BC are equivalent to
periodic BC but with parameters 
$J_{z}$, $t_{z}$, $t_{xz}$, and $t_{yz}$ 
reduced in half from their actual values. 
Remember also that $k_z$ can only be
$0$ or $\pi$. Thus, for a bilayer assumed
symmetric with respect to the $x\leftrightarrow y$ interchange in the
planes,
and spin symmetric ($\vartheta_{\alpha} =1$),
two branches of the magnon spectrum are found, namely
$$
\omega_{\bf q}=SJ\sqrt {16-\tau^2_{\bf q}+\frac{2J_{\perp}}{J}
(4\pm\tau_{\bf q})}
\eqno(6)
$$
\noindent
where ${\bf q}$ is a 2D vector, $J=J_x = J_y$ is the in-plane 
coupling, $J_{\perp}=J_z$
is the interlayer coupling, and
$\tau_{\bf q}=2(\cos q_{\bf x}+\cos q_{\bf y})$.\cite{Bonesteel}.

With this information,  we can find the quasiparticle hole 
spectrum in the AF background
which is defined by the position of the peak with the lowest binding
energy (assuming its intensity $Z$ is finite) in the spectral function
$$
A({\bf k},\omega)=-\frac{1}{\pi}Im\frac{1}
{\omega-\epsilon_{\bf k}-\Sigma({\bf k},
\omega+i\delta)},
\eqno(7)
$$
\noindent
where the self-energy $\Sigma({\bf k},\omega)$ is calculated
self-consistently from
$$
\Sigma({\bf k},\omega) =\frac{1}{N}\sum_{\bf q}\frac{M^{2}_{{\bf k},{\bf q}}}
{\omega-\omega_{\bf q}-\epsilon_{{\bf k}-{\bf q}}-\Sigma({\bf k}-{\bf q}, 
\omega-\omega_{\bf q})},
\eqno(8)
$$
which corresponds to the sum of magnon rainbow diagrams, as described in
previous literature.\cite{Rainbow} Eq.(8) allow us to consider, at least
in part, the dressing of the hole by spin excitations, and it has been
shown that the predictions arising from this approach are in excellent
agreement with Exact Diagonalization numerical results 
for 2D  finite clusters.\cite{Rainbow}
Thus, we are
confident that the rainbow approximation may maintain its accuracy when
applied to bilayer and 3D systems where numerical results are
difficult to obtain.
The solution of Eq.(8) is obtained numerically on finite but
large clusters.

While the SCBA formalism described here was setup for arbitrary values of
the many couplings appearing in Hamiltonian
Eq.(2), in the results reported in the following sections, 
we have neglected NNN hoppings in all
directions and set the magnetic anisotropy parameter $\vartheta_{\alpha}$
to 1(spin
isotropic system). 
We also assumed $xy$ symmetry in the planes, i.e. orthorhombic
distortions are not considered here. Thus, in the rest of the paper
 we will use
the following notations: $J$ and $t$ will refer to the in-plane exchange
and hopping
parameters, while
$J_{\perp}$ and $t_{\perp}$ correspond to the inter-plane ones.
The generality of Eqs.(2-8) would allow the interested
reader to obtain SCBA results for arbitrary situations where spin or
lattice anisotropies are important, or including NNN hopping amplitudes.

Regarding the numerical values of the couplings, we have introduced
 relations between exchange and hopping amplitudes to simplify the
multiparameter analysis. In particular, we have here assumed
$\frac{J_{\perp}}{J}\,=\,(\frac{t_{\perp}}{t})^2$, which arises from 
the $t-J$ model when derived from the one band Hubbard model in
strong coupling. In addition,
we have fixed $t\,=\,0.4$ eV throughout the paper. Thus, as parameters
we only have $J/t$ and $J_{\perp}/J$.

The actual value of the inter-plane exchange is controversial. The
two-magnon peak in Raman spectra experiments for ${\rm YBa_2Cu_3O_{6+x}}$ is
consistent with an in-plane $J\sim$
125 meV, with a negligible interplane exchange.\cite{Ratio} 
Recent 
measurements of the nuclear quadrupole and nuclear magnetic resonances
on Cu-sites in ${\rm
YBa_2Cu_3O_{15}}$ (alternating 1-2-3 and 1-2-4
structures) suggest a ratio
$\frac{J_{\perp}}{J}\,\sim\,$0.3.\cite{Ratio1} However, a different value
$\frac{J_{\perp}}{J}\,\sim\,$0.55 was found from infrared
transmission and reflection measurements, after applying 
linear spin wave theory.\cite{Ratio2}
A hopping amplitude ratio $\frac{t_{\perp}}{t}\,=\,$0.4 was reported
on the basis of first-principles linear 
density approximation calculations,\cite{Andersen}
corresponding to $\frac{J_{\perp}}{J}\,\sim\,$0.16 with our convention.
In view of these discrepancies, in most of the results 
below the perpendicular exchange is consider a
free parameter. 


\section{Results}

\subsection{Bilayer.}

Let us first discuss the one hole spectrum in the AF bilayer system
using the exchange ratio
$\frac{J_{\perp}}{J}\,=\,(\frac{t_{\perp}}{t})^2\,=\,0.16$ suggested
by band structure calculations, $t=0.4$eV and  
$J\,=\,0.125$ eV, as found in some experiments 
($J/t=0.3125$). The self-consistent calculation was
performed on a $16 \times 16 \times 2$ cluster.  In Fig.1 
the spectrum is presented. It has two
branches that correspond to the two possible momenta in the z-direction i.e.
$0$ and $\pi$. These branches obey the relation $\epsilon_{+}({\bf k})\,=\,
\epsilon_{-}({\bf k}+{\bf Q})$, where ${\bf Q}\,=\,(\pi, \pi)$ and
${\bf k}$ is the 2D momentum ($\epsilon_{\pm}({\bf k})$ here 
denote the even and odd branches of the spectrum). 
This relation is induced by the presence of
antiferromagnetic long-range order in the system, which effectively reduces
the size of the Brillouin zone.
The minima of the branches are close to, 
but not exactly at, ${\bf k}=({\pi/2, \pi/2})$ i.e.
where the minimum of the one-layer dispersion resides.
This small effect is observed from fits of the data using trigonometric
functions and it is barely noticeable in Fig.1,
but as we increase
$J_{\perp}$ the splitting of the minima away from ${\bf k}=({\pi/2,
\pi/2})$ becomes larger, and at 
$J_{\perp}\,=\,J$ it is clearly visible.
The bilayer one-hole spectrum 
can be accurately fit by NN, NNN and next to NNN
hoppings amplitudes, 
namely $\epsilon_{\pm}({\bf k})\,= \pm\,0.01(\cos k_x+\cos k_y)\,
+\,0.099\cos k_x\cos k_y\,+\,0.033(\cos 2k_x+\cos 2k_y)$ (eV) (see Fig.1).
Note that each individual branch contains a NN amplitude which is
nonzero. This is necessary since individually the even and odd branches
are not invariant under a momentum shift in $\bf Q$.
It should also be remarked that near $(\pi,0)$ there are
``flat'' dispersion regions,
as observed in previous studies of the $t-J$ 
model.\cite{flat-theory,other-flat} It
has been argued that AF correlations induce these features in models of
correlated electrons even when the $\xi_{AF}$ is only a couple of 
lattice spacings.\cite{moreo} Fig.1 shows that these features also survive the 
introduction of a realistic bilayer coupling.

\begin{figure}[htbp]
\centerline{\psfig{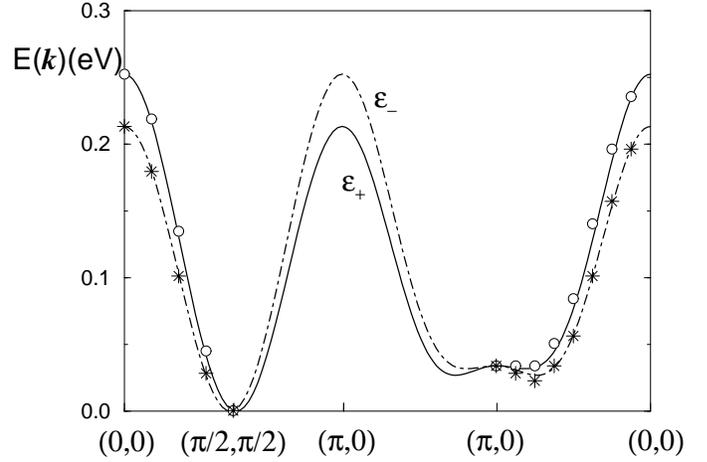}
}
\vspace{.2cm}
\caption{SCBA spectrum of one hole in an AF bilayer for parameters
$\frac{J_{\perp}}{J}\,=\,(\frac{t_{\perp}}{t})^2\,=\,0.16$,
$t=0.4$eV and  $J\,=\,0.125$ eV, on a
$16 \times 16 \times 2$ cluster.
Open circles belong to the 
bonding branch $\epsilon_{+}$, while
stars correspond to the antibonding branch $\epsilon_{-}$. 
Results between
$(\pi/2,\pi/2)$ and $(\pi,0)$ (not shown) are trivially related 
by symmetry to others presented in the same figure.
The solid (dot-dashed) line
is a fit using trigonometric functions
for the bonding (antibonding) branch.}
\end{figure}

\begin{figure}[htbp]
\centerline{\psfig{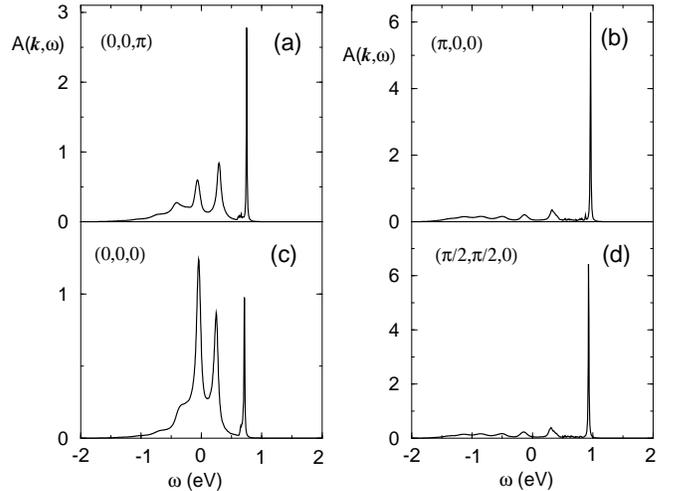}
}
\vspace{.2cm}
\caption{SCBA spectral functions of a hole in an AF  bilayer at high symmetry
points for the same values of parameters used in Fig.1, and with an artificial
 broadening
$\delta=0.1t$ for the $\delta$-functions. 
The momenta $(k_x, k_y, k_z)$ are indicated.
$k_z\,=\,0$ corresponds to the bonding branch, while
$k_z\,=\,\pi$ to the antibonding.}
\end{figure}

Our SCBA calculations indicate that the 
antiferromagnetic bilayer has quasiparticle excitations
 since a large peak appears at the bottom of the hole spectral
function $A({\bf k},\omega)$.
Typical spectral functions for the high symmetry points
are presented in Fig.2. 
The quasiparticle peak carries a substantial percentage of the spectral weight
in agreement with the
calculations for a single layer.\cite{review} Note also that
the closer the momentum to the
bottom of the band, the stronger the quasiparticle peak. We
observed a similar
behavior even for interlayer exchanges as large as
$\frac{J_{\perp}}{J}\,=\,3$.

The evolution of the Fermi Surface (FS) as the SCBA hole dispersion
is populated
in the rigid band picture is shown in Fig.3. At very low hole 
density it starts with hole pockets
around $({\pi/2, \pi/2})$ (Fig.3(a)) for the even and odd branches. 
The pockets are longer along the $(0,\pi)-(\pi,0)$ direction
than along the main diagonal in the Brillouin zone as observed in similar 
studies for the single layer problem.\cite{Trugman,flat-theory} As
these pockets grow in size, the Fermi level eventually hits the
saddle-points and the FS changes its topology
(Fig.3(b) and (c)), becoming a
large FS when the chemical potential is  above  the
saddle points (Fig.3(d)). 
It should be remarked that after the hole pockets disappear, the FS
acquire quasi-nesting features, which may lead to an enhancement
of certain susceptibilities. The
energy scale of the rapid change from hole pockets
to a large FS is about 400K. This implies that a 
strong temperature dependence should be expected in 
transport measurements of
a doped AF bilayer,
similar to those observed in a single AF layer.\cite{Trugman}
Actually the overall FS evolution obtained from the rigid band filling of the
bilayer dispersion for
$\frac{J_{\perp}}{J}\,=\,(\frac{t_{\perp}}{t})^2\,=\,0.16$ 
 is qualitatively similar to results found for just
one ${\rm CuO_2}$ plane using the same approximations,\cite{Trugman,flat-theory}
with the main difference being  the presence of two branches.
Identifying similar features in experiments could give us information
 about the strength of the bilayer coupling and antiferromagnetic correlations.

\begin{figure}[htbp]
\centerline{\psfig{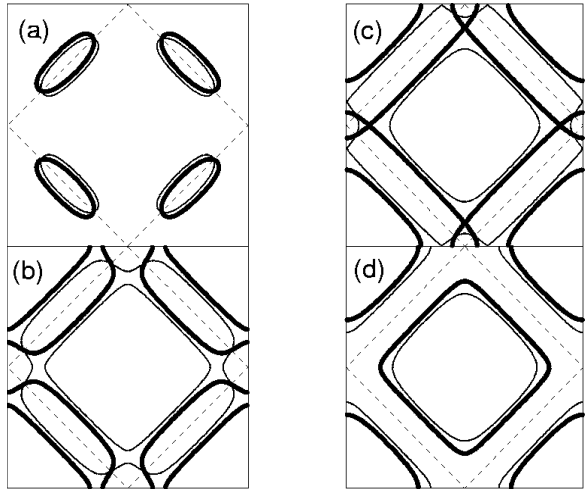}
}
\vspace{.2cm}
\caption{Evolution with doping of the Fermi Surface  obtained by a rigid band
filling of the one hole SCBA spectrum presented in Fig.1. The density grows
moving from (a) to (d). The thick (thin) line corresponds to the bonding
(antibonding) branch.}
\end{figure}

\subsection{3D Cubic Lattice}

In this subsection, the properties of a hole in
a 3D antiferromagnetic environment will be analyzed.
The SCBA quasiparticle hole dispersion for the case of an isotropic 
cubic lattice (i.e. $\frac{J_{\perp}}{J}=(\frac{t_{\perp}}{t})^2=$1) is
shown in Fig.4. The 
numerical self-consistent calculation was carried out on an $8 \times 8
\times 8 $
cluster with parameters chosen as 
$t=0.4$eV and $J=0.16$eV, which gives $\frac{J}{t}\,=\,0.4$.
The minimum
of the dispersion is at $({\pi/2, \pi/2, \pi/2})$, which is natural
based on the results reported for the one and two coupled planes.
Our best fit of the numerical data corresponds to the following
dispersion:
$\epsilon({\bf k})\,= 0.082(\cos k_x\cos k_y+
\cos k_y\cos k_z+\cos k_x\cos k_z)\,
+\,0.022(\cos 2k_x+\cos 2k_y+\cos 2k_z)$ (eV). As in the case of the one
layer system, holes tend to move within the same sublattice to avoid
distorting the AF background.

\begin{figure}[htbp]
\centerline{\psfig{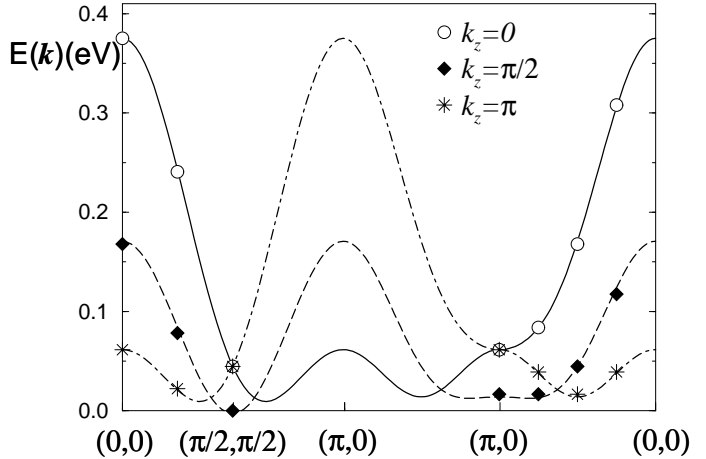}
}
\vspace{.2cm}
\caption{SCBA spectrum of one hole in an isotropic 3D AF with $J/t\,=\,0.4$.
The momentum in the z-direction is indicated.
The lines are fits to the numerical data:
the solid line corresponds to $k_z\,=\,0$, 
dashed to $k_z\,=\,\pi/2$, and
dotted-dashed to $k_z\,=\,\pi$.
Points between
$(\pi/2,\pi/2)$ and $(\pi,0)$ (not shown) are related 
by symmetry to others presented in the figure.}
\end{figure}

It is interesting to analyze the dependence of the quasiparticle
dispersion total bandwidth, $W$, with the ratio $\frac{J}{t}$ for the isotropic
3D AF. Here $W$ is defined as the difference in energy between the state with 
the highest energy, typically at $(0,0,0)$ and $(\pi,\pi,\pi)$, and the state
with the lowest energy at $({\pi/2, \pi/2, \pi/2})$.
Fig.5 shows $W$ as the ratio $\frac{J}{t}$
varies from 0.1 to 1.  Results for a single layer on
an $8 \times 8$ cluster
are also shown for comparison. The overall behavior of the
bandwidths is qualitatively the same in both cases, with $W$
for a 3D AF being slightly larger. In both cases $W/t$ grows approximately
linearly for small $J/t$, and reaches saturation around
$\frac{J}{t}\,\sim\,0.6$. 
In the linear regime, we find
$W \sim 2.5J$ showing that
the characteristic energy scale of the dispersion is $J$ rather than
$t$, a well-known result obtained before in the context of holes in
2D antiferromagnets.\cite{review} It is interesting to observe that 
the increase in
the dimensionality of the problem does not change dramatically the bandwidth of
holes. 

\begin{figure}[htbp]
\centerline{\psfig{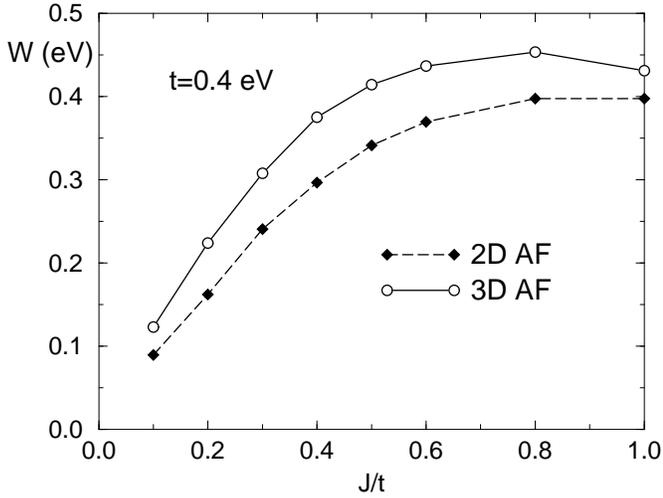}
}
\vspace{.2cm}
\caption{Open circles represent the  
bandwidth $W$ vs $J/t$ obtained from the SCBA quasiparticle hole spectrum
for an isotropic 3D AF. 
Filled diamonds are the bandwidth of an isotropic 2D AF (evaluated on an
$8 \times 8$
cluster) shown for comparison. Lines are 
drawn to guide the eye.}
\end{figure}



As for planes and bilayers, the SCBA calculations show a strong
quasiparticle peak in the hole spectral function $A({\bf k}, \omega)$  
of an isotropic 3D AF. In Fig.6 results are shown at particular high
symmetry points. The parameters
are the same as those used to calculate 
the dispersion of Fig.4. We checked that the relative
weight of the quasiparticle peak increases with the ratio
$\frac{J}{t}$.
The
quasiparticle weight $Z$ is similar in 2D and 3D systems, at least
at the bottom of the hole spectrum.

\begin{figure}[htbp]
\centerline{\psfig{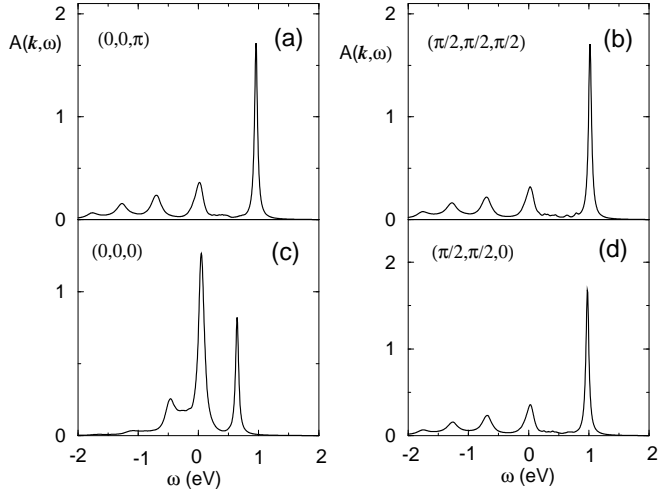}
}
\vspace{.2cm}
\caption{Spectral functions of a hole in an isotropic 3D AF at high symmetry
points for the same values of parameters used in Fig.4 with a broadening
$\delta=0.1t$ for the $\delta$-functions.
The 3D momenta ${\bf k}$ is indicated.}
\end{figure}

\section{Superconductivity.}

If there is a source of attraction between
 the quasiparticles described in the previous sections,
then the ground
state of the system could become superconducting.
In the underdoped regime, AF correlations are strong and, thus, it is natural
to study pairing mediated by AF fluctuations assuming
that the dispersion at half-filling is not drastically distorted when
a finite but small hole density is studied. 
Since $\xi_{AF}$ is likely 
of the order of only a couple of lattice spacing at realistic dopings,
a $real-space$ approach to pairing is more suitable than a picture where
extended states (magnons) are interchanged between carriers ($k-space$
approach). While these two variations of AF-mediated pairing are
quantitatively different, nevertheless it has been shown that 
they lead to the same symmetry for the superconducting
order parameter (SCOP).\cite{afvh} Thus,
independent of the discussion of real-space vs k-space approaches, we 
believe that it is safe to use the real-space pairing ideas of Ref.\cite{afvh}
to study the symmetry of the SCOP
for the case of a bilayer and 3D AF systems.

As explained before,
the simplified picture introduced in previous 
literature\cite{afvh} is to use the
spectrum of a hole, calculated numerically or using the SCBA, 
as the dispersion of the 
quasiparticles which interact through a NN attraction regulated by a
parameter $V$
(of the order of $J$). This interaction is motivated by AF-induced pairing
since it can be shown that a potential $V({\bf q})$ in momentum 
space maximized at ${\bf q=Q}$ and smeared by a finite $\xi_{AF}$, induces a
real-space potential which is dominated by a NN attraction. 
Previous estimations suggest a value $V=0.6J$ from
the study of the two-holes bound state on a single layer.\cite{afvh} Here,
the same value will be employed
for the in-plane NN interaction, while for the interaction in the 
z-direction $V_{\perp}=0.6J_{\perp}$ will be used.
In the absence of accurate Exact Diagonalization or Monte Carlo results for
the dispersion of holes away from half-filling, a rigid-band filling of the
SCBA quasiparticle spectra obtained at half-filling will be assumed
in the BCS gap equation.
Previous studies for single layer systems
have shown that this approach provides a SCOP with a symmetry
consistent with other more traditional methods, and in addition the actual
values of $T_c$ and its density dependence are in qualitative agreement
with experiments.\cite{afvh} 
Thus, it is natural to employ the same approach to
analyze the SC properties of doped bilayers and 3D antiferromagnets.

\subsection{Superconductivity in a bilayer.}

Using the bilayer SCBA hole dispersion and the hole-hole 
NN attraction described before, we have
studied superconductivity within the BCS formalism.
To calculate the SCBA hole dispersion,
we fixed $\frac{J}{t}\,=\,0.3125$ and allow for the interlayer coupling
to take the values $\frac{J_{\perp}}{J}\,=\,$0.16, 0.3,
0.5, 1, 2 and 3. Solving numerically the gap equation on large clusters,
we observed that 
for the first three
ratios $\frac{J_{\perp}}{J}$ the symmetry of the
superconducting order parameter (SCOP)
is $d_{x^2-y^2}$ for all the hole densities
where superconductivity exists. This is the region analytically
connected to the physics of one layer, where a similar condensate was
found in previous papers. In  Fig.7, we show $T_c$ vs  
the quasiparticle hole density 
for a bilayer with $\frac{J_{\perp}}{J}\,=\,$1. This
dependence is also shown for the case of
a single layer with the same $J/t$, for comparison.
Both curves exhibit the ``optimal doping'' behavior and high values
of  $T_c$ of the order of
100K. The difference in $T_c$ between a bilayer and a single
layer is small for practical purposes, except for the region of
high densities where
a bilayer changes the symmetry of the superconducting condensate into
an ``s-wave'' state to be 
discussed below (note that a quasiparticle weight $Z$ smaller than one
implies the existence of incoherent weight in the spectral function
 at energies larger than those
of the q.p. band. Then, ``high density'' in this context does not 
imply a very small electronic density $\langle n \rangle$).
Similar conclusions were obtained in Ref.\cite{Jose}
where an Exact Diagonalization study of coupled layers was carried out
to calculate the superconducting pairing correlations. 
No important  enhancement of pairing correlations 
changing from a single $t-J$ layer to a bilayer was reported.\cite{Jose}

\begin{figure}[htbp]
\centerline{\psfig{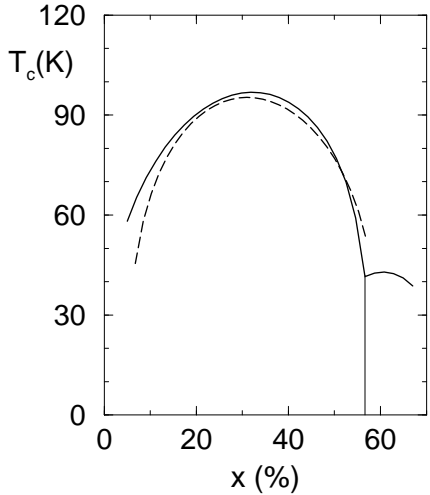}
}
\vspace{.2cm}
\caption{Superconducting critical temperature 
$T_c$ (solid line) vs quasiparticle hole density for an AF bilayer 
with $J/t\,=\,0.3125$
and $J_{\perp}/J\,=\,1$
(see text for details). The dashed line 
is $T_c$ vs quasiparticle hole density for a
2D AF plane with $J/t\,=\,0.3125$. 
Both lines are calculated within the framework
of the real-space pairing approach and the BCS gap equation. For bilayers, at
high q.p. densities ($x \sim 55\%$) there is a 
transition to an ``s-wave'' state.}
\end{figure}

According to
the classification of Liu, Levin and Maly\cite{Levin} the $d_{x^2-y^2}$
 state that we observed is labeled $(d,d)$,
which means that a gap with the symmetry $d_{x^2-y^2}$
opens in both the bonding and antibonding Fermi surfaces (i.e. the Fermi
surfaces corresponding to $k_z = 0$ and $\pi$ shown in Fig.1). 
This d-wave channel dominates up to values close to 
$J_{\perp}\,\sim\,J$. At the ratio $\frac{J_{\perp}}{J}\,=\,$1 the
symmetry of the SC order parameter changes only at 
high densities (i.e. around 55\% filling of the quasiparticle band as
shown in Fig.7) to a
mixture of $d_{3z^2-r^2}$ and $s_{x^2+y^2+z^2}$, with real
amplitudes of the same sign. 
$s_{x^2+y^2+z^2}$ is the 3D analog of the extended s-wave function.
The relative weight of the $d_{3z^2-r^2}$ component is dominant in the bilayer 
case. In general, the tetragonal point symmetry group consists of 10
irreducible representations among which only five correspond
to singlet pairing.
The functions ${3z^2-r^2}$, ${x^2+y^2+z^2}$  and also a $constant$
function 
have the same transformation properties with respect to the
tetragonal symmetry operations and thus belong to the same
representation $\Gamma_{1}^{+}$ of  the tetragonal lattice
symmetry group. Then, they mix together and they form 
the ``s-wave'' representation with respect to rotations in the planes,
while $d_{x^2-y^2}$ belongs to $\Gamma_{3}^{+}$ and $d_{xy}$ to
$\Gamma_{4}^{+}$ for a tetragonal system.\cite{Sigrist} Note that
in an orthorhombic lattice $d_{3z^2-r^2}$, $s_{x^2+y^2+z^2}$  as well as
$d_{x^2-y^2}$ 
correspond to the same
representation $\Gamma_{1}$. 

A less trivial issue is how the symmetry
evolves from a mixture of $d_{3z^2-r^2}$ and $s_{x^2+y^2+z^2}$ to
the so-called $d_z$ state\cite{dz} (the $d_z$ state is a singlet along
the link in the z-direction of the bilayer, i.e. it is ``s-wave'' with
respect to $\pi/2$ rotations of the planes).
This transition occurs as the ratio
$\frac{J_{\perp}}{J}$ grows at a fixed density.
The relative weight of 
$s_{x^2+y^2+z^2}$ in the mixture increases until both functions become
equally weighted with the same signs of the weight coefficients
in the limit of $J_{\perp} \gg J$. In this situation
the mixture is exactly equivalent to the $d_z$ symmetry. In our
calculations a coupling $\frac{J_{\perp}}{J}\,=\,$2 is enough to
have almost $d_z$ symmetry for all densities, with a substantial increase
of $T_c$ up to 150K. The mixture of $d_{3z^2-r^2}$ and $s_{x^2+y^2+z^2}$
can also be viewed as the mixture of $d_z$ and the extended in-plane s-wave
$s_{x^2+y^2}$, or in terms of odd-even gaps
$\Delta_{\pm}({\bf k})\,\propto\,\pm \alpha\,+\,\beta(\cos k_x+\cos k_y)$,
where ${\bf k}$ is the in-plane momentum of the quasiparticle
(the even and odd gaps differ by a constant corresponding to the 
inter-plane pairing amplitude).
In a similar fashion, as the ratio $\frac{J_{\perp}}{J}$ increases,
the relative weight of the in-plane extended s-wave  decreases. 

\begin{figure}[htbp]
\centerline{\psfig{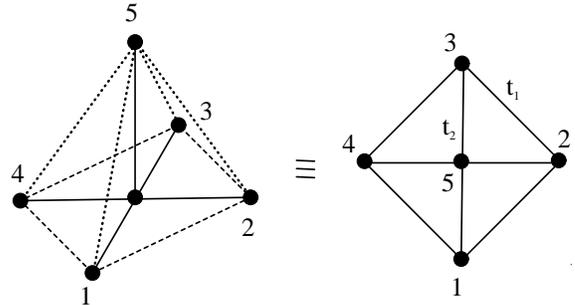}
}
\vspace{.7cm}
\caption{Schematic representation of the strong $V$ coupling limit of the
two quasiparticle problem for the case of a bilayer. The dashed and
dotted lines represent the effective hopping amplitudes contained in the
SCBA dispersion that we label $t_1$ and $t_2$, respectively. 
For more details see text.}

\end{figure}

There is a simple intuitive way to understand the evolution with
$J_{\perp}/J$ of the
symmetry of the superconducting order parameter found here.\cite{naza1}
Consider the limit of Eq.(1) applied to a bilayer
where the NN in-plane and inter-plane 
attractions dominate over the kinetic energy. 
In this limit, let us analyze the two
quasiparticle problem and its bound state. 
The quasiparticles will be roughly at a distance of one lattice
spacing (tight bound state). In the reference frame of one particle, the
other can occupy any of the five NN sites (four in the same plane and
one in the other plane). These sites are actually $linked$ by the
dispersion in the SCBA since the five sites belong to the same
sublattice. In other words, the wave function of one quasiparticle
orbiting around the other can be found in this limit from a simple
5-site 1-particle problem as shown schematically in Fig.8, with the
ratio of hopping
amplitudes $t_1$ and $t_2$ simulating the effect of $J_{\perp}/J$.
Solving this
trivial problem we find that for $J_{\perp}/J <1$ (or $t_2/t_1 <1$)
the lowest energy is obtained for a state with amplitudes 
$c,-c,c,-c,0$ $(c>0)$ at
sites $1,2,3,4,5$, respectively, which is the $d_{x^2 - y^2}$ state. 
On the other hand, for $J_{\perp}/J >1$ (or $t_2/t_1 >1$) the ground
state amplitudes have the same value (with its sign) in-plane, and a different
value and sign
out of the plane, as expected from the discussion above. Note also that
these simple calculations illustrate the fact that a NN attraction and
on-site repulsion is $not$ enough to have a tendency to form d-wave
pairs in a $dilute$ gas. In the present case this potential must be
supplemented by a particular hole dispersion that allows for the
movement of carriers within the same sublattice.

\subsection{Comparison with other theories}

In the literature, several types of interactions between quasiparticles
in a bilayer
have been proposed. Of particular interest  for our discussion
 are those based
on phenomenological magnetic susceptibilities in the presence
of dynamical short-range AF order, both in and between the planes.
Recently it has been suggested\cite{Mazin} that the SCOP in a bilayer
can be of a particular s-wave type named $(s,-s)$
with opposite signs in the odd and
even branches of the spectrum of carriers.
The assumption made in Ref.\cite{Mazin}
 is that the in-plane magnetic susceptibility 
$\chi_{\|}({\bf q})$ has the opposite sign, but the same absolute
value, as the
inter-plane susceptibility $\chi_{\perp}({\bf q})$. In other words,
$\chi_{\|}({\bf q})\,=\,-\chi_{\perp}({\bf q})\,=\,
\chi_{0}({\bf q})$, where $\chi_{0}({\bf q})$ is the phenomenological
susceptibility of a doped 2D antiferromagnet peaked at ${\bf q}\,=\,{\bf
Q}\,=\,(\pi,\pi)$
introduced to fit NMR data.\cite{MMP}
This assumption needs further justification since
the analysis of experimental results provides an estimation
5 meV $<\,J_{\perp}\,<$ 20 meV\cite{Monien} for the interplane coupling, 
while $J\,\approx\,$120 meV.
Thus, it is somewhat risky to assume that the spin correlations between
planes have the same strength as in the planes.
Then, it seems more natural 
to employ the RPA approximation for the
susceptibility\cite{Monien} given by
$\chi^{-1}_{\pm}({\bf q})\,=\,\chi^{-1}_{0}({\bf q})\mp J_{\perp}$,
where $\chi_{\pm}({\bf q})$ are the even and odd branches of the 
bilayer susceptibility, respectively.
In this case, $\chi_{\|}({\bf q})$ and $\chi_{\perp}({\bf q})$
are quite different in absolute value, i.e.
$|\chi_{\|}({\bf q})|\,\gg\,|\chi_{\perp}({\bf q})|$ in order 
to satisfy the applicability of the RPA approximation. 
We have checked by a direct
calculation, that this approach provides $d_{x^2-y^2}$ symmetry for the
SCOP on both branches of the spectrum i.e. the so-called 
$(d,d)$ state. This is 
in agreement with 
Ref.\cite{Levin} where a similar phenomenological in-plane
susceptibility derived from the neutron scattering data was used to
construct the effective potential with an arbitrary relation between
the in-plane and inter-plane coefficients. 
The above mentioned RPA model is certainly relevant to the
more phenomenological approach used in this paper
if one considers the expansion of 
$\chi_{0}({\bf q})$ in real space, assuming that the 
effective coupling between quasiparticles is proportional to $\chi_0$.
For correlation lengths $\xi\sim2-3$ lattice spacings,
the expansion consists of a strong on-site repulsion $U$, and also a relatively
strong NN attraction $V$ in-plane, with a ratio 
$\frac{U}{|V|}\sim3$. The NNN interaction is repulsive and about
one order of magnitude weaker than $U$.
Other terms are negligibly small.
Then, the phenomenological interaction 
used here and in previous literature\cite{afvh}
is very closely related to the RPA interaction in the realistic regime 
$J_{\perp}\,\ll\,J$. To show this result, 
first let us expand the susceptibility as
$$
\chi{\pm}({\bf q})\,=\,\frac{1}{\chi^{-1}_{0}({\bf q})\pm J_{\perp}}
\simeq \chi_{0}({\bf q})\mp J_{\perp}\chi^{2}_{0}({\bf q}).
\eqno(9)
$$

\noindent Using a 3D notation one can write $\chi({\bf q}, q_z)
\simeq \chi_{0}({\bf q})-\cos (q_z) J_{\perp}\chi^{2}_{0}({\bf q})$,
for the case of a bilayer with $q_z\,=\,$0 or $\pi$. The 
effective in-plane potential
must contain an strong on-site repulsion. Consequently,
$\chi_{0}({\bf q})$ can be written as $\chi_{0}({\bf q})\,=\,
u+f({\bf q})$, where $u$ is a constant proportional to the 
on-site repulsion $U$, and the function
$f({\bf q})$ satisfies the condition $\sum_{\bf q}f({\bf q})=0$.
Hence, the susceptibility can be rewritten as
$\chi({\bf q}, q_z,)
\simeq u+f({\bf q})-\cos (q_z) J_{\perp}[u+f({\bf q})]^2$. Assuming
that the ratio $\frac{f({\bf q})}{u}$ is small (i.e. the on-site
repulsion is strong), we finally get:
$$
\chi({\bf q}, q_z)
\simeq u+f({\bf q})- J_{\perp}u\,\cos (q_z)
- 2J_{\perp}u\,\cos (q_z)f({\bf q}).
\eqno(10)
$$

The first three terms are accounted for in the calculation discussed in
this paper both for a
bilayer and a 3D antiferromagnet since, as it was argued above, 
the largest contribution to $f({\bf q})$ 
comes  from the NN attraction. The last
term in Eq.(10) is relatively small compared to the other three, 
but it is assumed to have a significant contribution
in the models of Refs.\cite{Levin,Mazin} explaining the difference
between their results (i.e. $(s,-s)$) and ours
(i.e. mixture of $d_{3z^2-r^2}$ and $s_{x^2+y^2+z^2}$).
Nevertheless,  both cases
are ``s-wave'' in-plane and thus there is
no qualitative difference in their transformation properties (they 
belong to the same representation). In
addition, since
$J_{\perp}/J \ll 1$, both are unlikely to be stable in the
cuprates.




\subsection{Superconductivity in the isotropic 3D AF.}

To calculate $T_c$ and the symmetry of the
SCOP in the case of a three dimensional
isotropic AF, we here use the 
SCBA 
hole dispersion obtained at $J/t\,=\,0.4$ and again the BCS gap equation. As
expected,\cite{Miyake,Scalapino}
when $\frac{J_{\perp}}{J}\,=\,1$, the two channels
$d_{x^2-y^2}$ and $d_{3z^2-r^2}$ are degenerate, i.e. $T_c$
is the same for both and varies with the hole density
as it is shown in Fig.9. The critical temperature in 3D is
slightly smaller than for planes and bilayers.
The ``optimal doping'' behavior is also present in 3D, but it is
less pronounced than in 2D.



\begin{figure}[htbp]
\centerline{\psfig{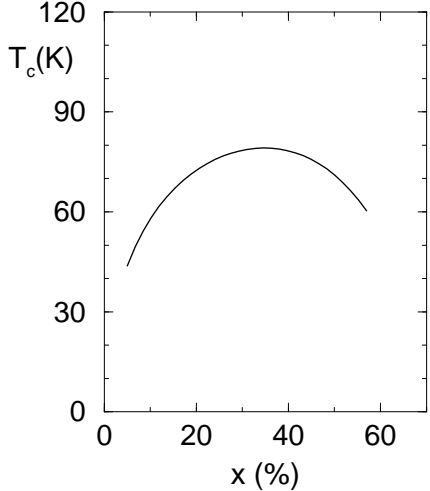}
}
\vspace{.2cm}
\caption{Superconducting critical 
temperature $T_c$ vs quasiparticle hole density for an isotropic
3D AF with $J/t\,=\,0.4$, using the BCS gap equation and Hamiltonian Eq(1).}
\end{figure}

When $\frac{J_{\perp}}{J}\,\neq\,1$, the system acquires tetragonal
symmetry and the channels $d_{x^2-y^2}$ and $d_{3z^2-r^2}$ are 
no longer degenerate. In fact, if $\frac{J_{\perp}}{J}\,<\,1$,
the gap possesses $d_{x^2-y^2}$ character, while if
$\frac{J_{\perp}}{J}\,>\,1$, it becomes again a
mixture of $d_{3z^2-r^2}$ and $s_{x^2+y^2+z^2}$, as it was
mentioned in the case of a bilayer. Again, as the ratio
$\frac{J_{\perp}}{J}$ increases, 
the symmetry of the SCOP
becomes $d_z$.  All these results can also be obtained following a
discussion similar to that of Fig.8 but now using six sites instead of five.

\section{Conclusions}

In this paper, we have investigated the properties of bilayer and 3D
antiferromagnets specially regarding the behavior of holes in such
environments and the eventual formation of a superconducting state upon
doping. The main assumption is that AF correlations remain of at least
a few lattice spacings in range
which, as a first approximation, 
allow us to use the dispersion of holes at half-filling and a
density-density NN attraction between quasiparticles also generated by 
antiferromagnetism.
Within the BCS gap equation
superconductivity is investigated. A small coupling in the z-direction
does not alter qualitatively the $d_{x^2 - y^2}$ superconducting state
found for a pure single layer plane. Only for large values of the
exchange in the direction perpendicular to the planes is that a
competing ``s-wave'' superconducting state becomes dominant. 

\section{Acknowledgments}

A. N. acknowledges the  support by the
National High Magnetic Field Laboratory, Tallahassee, Florida.
E. D. is supported by the 
NSF grant DMR-9520776. We thank
MARTECH for additional support.
\medskip


\end{document}